# 能源系统通证经济学：概念、功能与应用

张俊 [1, 2]　　王飞跃 [2, 3]　　陈思远 [1]

摘要：传统的中心化的能源系统具有管理困难、激励不足的缺点，而区块链技术具有去中心化、自治性、透明不可篡改、可追溯性等特性，能有效地应用在能源系统中。结合通证经济学和区块链技术的能源系统通证经济体系，具有强激励、精细化、低成本等特点，能够极大地调动清洁能源和用户的积极性，解决了大量弃风弃光、传统需求侧响应项目中响应迟缓等问题，为提升能源系统高效运行、节能减排提供了保障。本文先对通证经济体系的概念和功能进行阐述，然后对通证经济体系在能源系统中应用的可行性进行了分析，最后提出了通证经济体系的一种应用模式，并利用综合能源系统进行了示例说明。

关键词：能源系统，通证经济，清洁能源，需求侧响应，节能减排

# Token Economics in Energy Systems: Concept, Functionality and Applications

ZHANG Jun[1, 2]　　WANG Fei-Yue[2]　　CHEN Si-Yuan[1]

**Abstract** Traditional centralized energy systems have the disadvantages of difficult management and insufficient incentives. Blockchain is an emerging technology, which can be utilized in energy systems to enhance their management and control. Integrating token economy and blockchain technology, token economic systems in energy possess the characteristics of strong incentives and low cost, facilitating integrating renewable energy and demand side management, and providing guarantees for improving energy efficiency and reducing emission. This article describes the concept and functionality of token economics, and then analyzes the feasibility of applying token economics in the energy systems, and finally discuss the applications of token economics with an example in integrated energy systems.

## 1.引言

### 1.1 区块链与通证经济

比特币和以太坊的横空出世让越来越多的人开始关注其底层的区块链技术[1]。区块链，是一种可以实现全网共识共同维护且保存所有历史交易数据的分布式数据库，其采用的时间戳、非对称加密、分布式共识、可灵活编程等技术让去中心化、高可信度的自治系统成为可能[2]。从区块链1.0的比特币、到区块链2.0的以太坊、再到如今区块链3.0[3]的综合应用，区块链颠覆了传统互联网的运行模式，使越来越多的人相信区块链就是未来的互联网新形态。

然而，区块链终究只是一种底层技术，其技术价值并不一定能创造足够大的社会

价值,真正能让区块链实现价值转换的,是具有改变生产力和生产关系的通证与通证经济体系。围绕"通证"进行设计和构建的区块链系统将会是未来趋势。实际上,比特币可以视作第一个"通证经济"模型。通证经济体系(Token Economy)是先于区块链发展而出现的一个概念,早在19世纪初期,英国耶鲁大学教授 Alan E. Kazdin 就将通证经济体系用于教育与治疗领域[4]。本文中定义的通证经济学及其体系是通过激励机制,构建了一种全新的生产关系,再配合区块链上高可信度的大规模协作模式,实现了大规模的自组织生态系统,促使生态系统中的每个参与者形成共识、共存的协作关系。通证之所以称为区块链最具特色的应用,主要源于以下几个方面:1)区块链是个天然的密码学基础设施,在区块链上发行和流转通证就是密码学意义上的安全可信;2)区块链是一个交易和流转的基础设施,而通证所需的具有高流动性和快速交易性的环境就是区块链的一个根本的能力;3)通证必须有内在价值和使用价值,而区块链通过智能合约可以为通证赋予动态的用途。

值得注意的是,通证经济体系和区块链其实是可以独立存在的,但是没有了"通证经济"激励的"区块链"只能是一个分布存储数据库,而没有了"区块链"支撑的"通证经济"也不能实现权益的快速流动与安全交换,通证经济体系与区块链技术是毫无疑问的最佳搭档。

## 1.2 通证经济学概述

通证经济学是一个利用干涉手段去提升行为改变的系统方法,经过数年的演变,这个系统对很多环境、行为和个人都十分有效。在经济学理论中,通证经济体系是一个与增强(reinforcement)有关的复杂系统,增强物(reinforcer)可以作为交换的媒介用作购买物资、服务、权限等。增强理论[5](Reinforcement Theory)是通证经济体系的理论基础,在1971年由 Skinners 提出,该理论认为某种行为的产生受某种结果的影响,故适当的奖励可能左右他人的行为。通证经济体系正是利用这一原理,用一种本来不具有增强作用的物体(如电力货币等)为表征,让它与具有增强作用的其他刺激物(如食品、服务、美元等)相联结,让这一种表征物变成具有增强力量的东西。这一种经由制约过程而获得增强力量的表征物通常被称为制约增强物(Conditioned Reinforcer),而能够积累并可兑换其他后援增强物(Backup Reinforcer)的制约增强物就称为通证(token)。通证经济体系的实施目标是通过经济制约手段增加实施者的良好期望行为,例如用奖励比特币的方式让区块链上的矿工做真实的账本,故建立通证经济体系的第一个过程是确定在实施过程中需要强化的期望行为。在形成期望目标之后,实施者需要确定一种通证,以及能与通证相配合的后援增强物(美元、物资等),并建立通证的兑换比率(例如,比特币与美元的兑换率)和量化每个目标行为(例如,矿工做成一笔账单奖励多少比特币)。当一个通证经济体系运行到成熟,实施者的期望目标将会移植到自然情境中,使自然社会的发展符合正确的导向。

## 1.3 能源系统概述

现阶段,人类社会对能源需求量的急剧增加导致了化石能源的过快和过度开发,并由此引发了人类社会对于环境污染和未来能源供应链的担忧。目前,我国分布式能源发展迅速,据不完全统计,早在2015年年底国内分布式能源总装机容量就达到了1100万kW,进入全球分布式能源大国行列。但是近几年来,分布式能源主要以天然气为主,风能、光伏等能源由于其不确定性遭到大量弃用,这一问题为未来能源发展带来深深的隐患。传统能源系统中,运行调度通常只针对单一的能源,不同能源系统相对独立,导致整体能源利用效率低。打破各能源体系之间的壁垒,构建集成互补的综合能源体系,实现多种资源的协调优化利用成为能源领域的主要趋势[6]。

为了提高能源利用率,国内外学者进行了多种能源系统的构思,目前较为主流的有综合能源系统与能源互联网。综合能源系统(Integrated Energy System, IES)作为热点研

究领域,在国内外已提出已久,它将电、冷、气、热、可再生能源等多种能源有机耦合,形成的不同能源形式联合运行优化,实现了能量的梯次利用与存储转换的能耗模式[7]。能源互联网(Energy Internet, EI)的概念是源于美国学者的著作《第三次工业革命》,希望借助智能通讯技术和智能电网,将各类分布式发电设备、储能系统和可控负荷有机综合,从而为用户提供清洁便利的能源供应,并使用户可以参与到能源的生产、消费与优化的全过程[8]。另外,在能源互联网的背景下,瑞士苏黎世联邦理工学院的 G.Anderson 教授提出了能源集线器(Energy Hub)模型的概念,它由转换模型 、分配模型和存储模型构成,可以实现电、冷、热的转换、分配和存储[9]。

综合能源系统和能源互联网虽然在一定程度上提高了能源系统的能源利用率,但是仍然没能解决分布式能源大规模上网供电存在的种种障碍。目前的能源系统主要以火电和燃气为主,分布式能源只能少量上网进行消纳,大多数分布式能源依靠国家补贴在运作,失去了能源系统建立的初衷。能源系统的构建与运行,不能仅仅靠供给侧的规划,用户的参与也是十分重要的一环。现阶段,综合能源系统和能源互联网虽然也存在需求侧响应的概念,但是用户与供给侧不能处于统一平台进行交易,交易的可靠性得不到保障,简单的激励措施不能调动用户的积极性。因此,建立一个自治可信的统一平台,让用户与供给侧以同等身份进行自由交易,是解决能源系统中各种问题的有效途径。

## 2 通证经济体系的结构与优势

### 2.1 通证经济体系的结构与功能

通证经济体系作为一个被广泛应用的工具,在不同领域中存在着很多细节差异,但是正常运转的通证经济体系都具有较为类似的核心结构。一般来说,一个有效的通证经济体系主要包括六个部分[10]:1)期望目标行为;2)一个可以作为制约(或通用)增强物的通证;3)后援增强物;4)通证的获得方式;5)通证与后援增强物的兑换方式;6)通证与后援增强物的兑换率。从十九世纪早期到现在,通证经济体系一直被用在教育和治疗领域[11],例如,在英国第一次工业革命时期鼓励学生通过挣取分数来换取自己喜爱的奖品,用激励的方法帮助精神病人减少不正常行为等。1965 年,人们尝试了一个与通证经济体系相关的实验项目[12],给与八个参与者在工作初期一个选择他们心中高优先或者低优先工作的机会,通证的增强作用通过 ABA 倒反设计进行评估(ABA 倒反设计分为三阶段,第一阶段和第三阶段为自然情况,第二阶段加入让人们选择低优先工作的增强物)。在第一个和第三个阶段,参与者挣取通证去选择他们心中高优先的工作,但在第二个阶段,参与者则挣取通证去选择心中低优先的工作。这个结果表明,参与者选择何种工作与通证和增强物有关,而忽略了自身的初始偏好,虽然这一结论的可信度在现在的标准中需要谨慎考虑,但它初步说明了通证经济体系在行为控制与规范方面的强大功能。

### 2.2 通证经济体系的优势

相比于其他基于增强理论的行为改变系统,通证经济体系有很多独特的优势,本节结合通证经济体系的结构特点进行详细描述。

(1)通证一般会选用具有安全、高效、流通速度快的物品(如虚拟货币),往往只需要很短的时间和很少的资源就能在系统中进行流通。相比较而言,其他后援增强物可能会受到环境约束的制约而产生传输延时等,不满足通证的高流通速度的要求,且安全性也难以保证。例如,现在的卡、券、积分、票等物品只能用作抵扣,不具备流通功能,其价值的安全性和可信度也得不到保障。

(2)通证一般不受激励的短暂性状态的影响,可以在多种条件下充当增强物。一个通证可以与多种后援增强物建立配对关系,尽管通证的增强作用也会随着后援增强物机制的波动而波动,但通证的存在会使这种波动对激励的影响降到最低。另外,由于通证的告诉流转和交易,每一个通证的价格都将在市场获得迅速的确定,这是价格信号

灵敏度的不及之处。

（3）通证经济体系的应用可以围绕着智能合约而展开。智能合约存在着千姿百态的创新，利用智能合约，任何人、任何组织、任何机构都可以基于自己的资源和负荷发行权益，与这些权益对应的通证是随时可验证、可追溯、可交换的，这种可信性和可靠性是传统金融手段无法实现的。

## 3 通证经济体系在能源系统中的应用模式

### 3.1 能源系统的需求和趋势

区域能源系统被认为是未来人类社会能源的主要承载形式，现在也有多个区域已经建立起多种形式的区域能源系统，但是世界范围内的能源系统均仅终端入手，例如欧盟初期的微网系统、美国的综合能源系统-冷热电联供(IES-CCHP)系统等。完整的能源系统涉及人类社会能源产生、输送、分配、消费环节，由于能量流以及相互之间交互的复杂性，对其整体开展研究难度极大。虽然信息物理能源系统(Cyber-physic Energy System，CPES)[13]和经济生态系统(Eco-economic System)可以为能源系统未来建模提供研发工具，但由于能源系统的复杂度上升，系统数据与能量流难免变得混乱，以区域综合能源系统为例，如图1为综合能源系统中的简易能量流结构图。

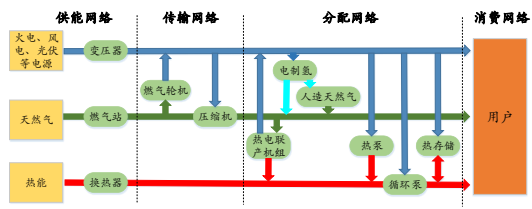

图1 IES 能量流结构图

能源系统在实现能源高效利用的同时，还需要兼顾分布式能源消纳与减少碳排放等任务，下面主要探讨能源系统中存在的三大需求：

（1）能源系统的激励问题

由于能源系统中能源种类多、能流密度大，为了保持能源系统的运行稳定，往往需要给与一定激励以使发电企业和用户按照管理预期目标来进行生产和消费，但是出于经济性考虑，管理者提供的激励往往并不丰富，甚至不能吸引都发电企业和用户。例如，当能源系统的输电网中存在阻塞时，管理者会对错峰用电的用户进行一定的奖励，但奖励的金额往往不会超过阻塞成本，并且对于一些生产型用户，停电的损失成本过于巨大，即使给与一定的费用补偿也不会削减负荷来缓解阻塞情况。

（2）新能源消纳问题

为了有效发展低碳电力，必须尽可能地让清洁能源上网供电，但是能源系统的特殊性与清洁能源的不确定性，往往会极大阻碍消纳。例如，在 IES 中，当热负荷需求上升时，CCHP 机组由于热电比调节能力有限，电能出力会随着热出力增多，从而使风电、光伏等新能源的上网空间被挤占，导致严重的弃风、弃光。另外，用户在使用新能源供电时牺牲了部分可靠性要求，为了鼓励用户使用清洁能源或者适当减少热负荷，需要对用户进行一定的补贴。但从能源生产到消费，很难确定用户群体的能量消费到底是来自天然气和分布式能源还是电网供电，换言之，无法确定哪些用户的行为为新能源消纳作出了贡献，也无法进行精准地奖励。

（3）能量流时间戳问题

能源系统中存在天然气网、热网和电网，在进行综合供能时，各种能源从生产、传输和使用的过程中均存在不同的延时，即无法得知用户此刻使用的热能和电能是什么时候生产的，也无法确定此时系统中的潮流情况，若发生潮流越限，可能会影响用户的用能需求。延时问题不仅增加了用户的不信任感，使用户与发电企业之间合约的可信度可可靠性下降，也给能源系统的潮流稳定性带来了一定的挑战。

针对能源系统存在的三大需求，目前很多学者在如何建立完善的激励机制或者调度方案上作了大量的工作，文献[14]提出了一种基于碳交易机制的电-气互联综合能源系统低碳经济运行，其中交易碳排放量可以使发电企业获得额外的利润；文献[15]提出了考虑多场景随机规划的冷热电联合系统

协同优化，其中也考虑了新能源消纳的问题。也有很多学者针对综合能源系统内的阻塞问题，对综合能源系统的稳定问题进行了研究，文献[16]对典型的区域综合能源系统的稳态问题进行了分析研究。上述这些工作虽然在节能减排、促进新能源消纳以及综合能源系统稳定性上提供了一些方法和思路，但是均只从系统规划和调度的角度出发，未能让用户参与其中。用户作为终端需求，是整个综合能源系统中十分重要的一环，不能激励用户进行需求侧管理，很难达到节能减排、减少阻塞的目的。传统的激励方式是分别在综合能源系统的源头和终端进行一定的激励，对清洁能源发电和需求侧响应用户提供补贴，对碳排放量大的发电厂和耗能过多的用户进行一定的惩罚，但受到经济因素制约，补贴和惩罚一般较为微薄，并不能调动发电企业和用户积极性，节能减排等策略难以实施。

## 3.2 通证经济体系在能源系统中应用的可行性分析

能源系统中能源种类多且分布较广，各类负荷错综复杂，能源与负荷交互过程较为复杂，人为地进行能源的生产与消耗的管理、调度与追踪是十分困难的。区块链技术利用其分布式、去中心化、自治化、可追溯性等特点呢，可以分别建立集中具有不同功能的区块链，即数据区块链、资产区块链、分析区块链、运营区块链以及支付区块链。利用数据区块链、分析区块链可以清晰地对能源流进行管理与追溯，利用运营区块链和支付区块链可以对能源的运输与消费进行精细化管控，区块链技术可以完美的处理能源系统中存在的传统方法无法解决的难题。文献[2]对区块链的构架与分层功能已作详细描述，本节仅在此基础上重点讨论在运营区块链和支付区块链上广泛应用的通证经济体系。

通证经济体系是利用制约增强物来增加实施者良好期望行为的手段，在综合能源系统中一般会选用带有经济效益的物体作为制约增强物，通过制约增强物的增强作为提高供能单位与用户的积极性。常规的增强物一般为货币补偿，例如对节能减排发电企业进行金钱奖励，但是当奖励达不到发电企业的预期时，这种措施往往难以继续执行。现阶段有些地区开展了碳交易市场，发电公司将自己剩余的碳排放量出售给其他公司，但是由于交易机制的复杂性以及交易对象必须实时匹配，目前取得的效果尚不明显。通证是可以与多种后援增强物进行交换的制约增强物，它可以兑换为货币抵用电费，也可以兑换为一些特殊权益（如优先发电权、优先买电权、能源通道使用权等），当发电企业和用户不满于完成目标行为（节能减排、缓解阻塞等）所带来的利润或获得的利润低于成本时，也可以选择用通证去换取一定的权益。通证与直接奖励的相比的优势在于，通证对应着多种后援增强物，它可以平抑一种或多种后援增强物的价值波动带来的影响，用户与发电企业可以选择自己想要的权益。

另外，在运营区块链和通证经济体系下，能源系统内任何角色都会存在多重身份，用户也可以作为资源提供者参与交易。用户在使用清洁能源的同时，也会作为减少碳排放的贡献者获得一定的通证；用户也可以采用负荷削减和转移，提供需求侧响应减少碳排放来获得通证。以前的需求侧响应项目之所以难以实施，一方面是激励不足，还有一方面是用户只能作为接受者被动地参与新能源消纳等项目。然而在运营区块链上，用户可以主动选择自己想买使用的能源类型，也可以主动选择对自己有利的规划项目，加上通证的增强作用可以满足用户的更多需求，用户的积极性被极大地激发了。再者，通证经济体系对与目标行为背道而驰的发电企业和用户也收取一定的通证，当发电企业或者用户的通证数量为负时，会限制其上网供电或购买电量，这既可以使发电企业和用户减少与目标行为相悖，也在一定程度上促进了通证在整个系统中的流通。

最后，正如上一节所述，能源系统中的能量流都有一定的延时，能量流时间戳的不准确性导致了用户供能的混乱。区块链中的数据层、网络层、共识层、合约层能实现生产、运输和消费的强追溯性，并通过实时数

据分析与实时数据共享，减少节点间数据共享的延时性。例如，让用户与发电企业签订智能合约，发电企业对用户供电时就会向用户发出数据，用户利用智能电表装置在区块链上获取这些数据从而确定自己所用的能源类型，这不仅增加了用户与发电企业之间智能合约的可信度，也减少了能源系统发生潮流越限的风险。

综上所述，通证经济体系在能源系统中的应用不仅解决一些传统的问题，还能增加能源系统的可靠性与稳定性，为能源系统的高效稳定运行提供了一个非常理想的理论和应用基础。

### 3.3 能源系统中通证经济体系的建立

#### 3.3.1 通证的获取方式

从本文所分析能源系统的需求出发，能源系统主要有节能减排和维持系统稳定的目的，节能减排可以通过减少碳排放量来体现[17]，而维持系统稳定主要表现在满足能源系统的运行约束即潮流不越限，因此考虑以碳排放贡献因子和阻塞贡献因子来量化和评估发电企业和用户的行为。以碳排放贡献因子为例，当发电企业的碳排放量未超过允许排放标准，按照剩余排放量所占总体允许排放量的比例计算其碳排放贡献因子（值大于 0），当碳排放贡献因子累计到一定程度（超过一个上阈值），奖励一定数量的通证；但若其碳排放量超过了允许排放标准，则需要按照超出排放量的大小计算其碳排放贡献因子（值小于 0），以同样的标准设置一个下阈值，收取一定数量的通证。同样地，根据用户使用清洁能源供能或者削减负荷的功率来计算其碳排放贡献因子，碳排放因子累计到一定数量则奖励一定数量的通证。对于阻塞贡献因子，只有在系统发生阻塞，发电企业或者用户减少输送和消费缓解阻塞才能获得一定的通证，阻塞贡献因子的计算与削减的功率大小相关，设置一个阶梯型的功率-阻塞贡献因子函数，对累计阻塞贡献因子达到一个阈值时奖励一定数量的通证。

#### 3.3.2 通证经济体系的构建

能源系统通证经济体系(Energy System Token Economy, ESTE)的建立必须依赖区块链群环境，区块链群中各个区块链是模块化关系，相互之间会提供所需数据及其接口，引用文献[2]中各区块链之间的信息交互关系如图 2 所示，其中具体细节本文将不再赘述。

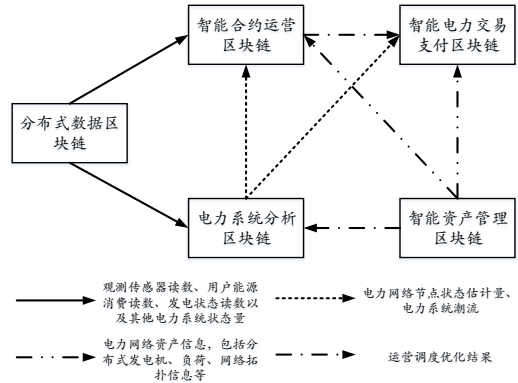

图 2 区块链信息交互示意图

ESTE 作为一个复杂的动态系统，它的建立必须从空间、时间和行为这三个方面进行建立，并且不能割裂这三个方面的关系，必须同时考虑三个方面的复杂性。

（1）空间维度构建

ESTE 的运行虽然不太依赖于实际的物理拓扑结构，但是考虑到综合能源系统中多能耦合的特殊性，发电企业和用户在进行交易时，必须满足潮流约束，因此必须在运营区块链上建立与实际物理实体（如发电机、燃气轮机、负荷、储能电池等）相对应的节点集，且根据能源的空间分布与耦合情况建立节点之间不同的连接关系（如天然气网、热网、电网等），最终形成一个虚拟系统。

（2）时间维度构建

潮流分析和状态估计是能源系统运营和管理的基础，通证的获取与高速流通都必须依靠可靠实时的潮流结果，传统的潮流分析与状态估计的方法已不能胜任 ESTE 的应用需求。ESTE 利用电力系统分析区块链对综合能源系统运行情况进行实时分析，且对能量的生产、运输、分配、消费过程都加以时间戳，将这些数据传输至运营区块链和支付区块链中。利用发电企业和用户在 ESTE 上签订的智能合约时间戳，与系统实时潮流状态相结合，智能合约主体在能源系统中的

生产和消费行为都具有追溯性，时间维度让能源系统中各类交易能在时间轴上形成链，为通证经济体系的实施创造了良好的环境。

（3）行为维度构建

在建立结构完整的虚拟系统之后，运营区块链会读取分析区块链的数据，对区块链上每一个用户（能源系统的发电企业和用户终端）的状态数据、能源消费数据、发电状态数据以及其他电力状态量进行读取和分析。这些数据都带有属性和时间戳，用户在运营区块链上签订的智能合约是完全可信的，因此发电企业和用户在能源在生产、运输、分配、消费的过程中占用的公共资源比例，以及对目标行为的完成情况，都可以较为容易的计算出来，也作为 ESTE 中奖惩的根据。

### 3.3.3 通证的激励方式

从三个维度构建的 ESTE 使整个系统中的能源全过程变得清晰透明，分析区块链会将分析数据传输给运营区块链和支付区块链，运营区块链会将所有智能合约的履行情况数据都传输给支付区块链，在支付区块链中有两类节点，第一类是计算节点（负荷潮流跟踪问题），另一类是交易节点（负责点对点资金的支付与收取）。根据潮流追踪计算用户的贡献因子，支付区块链会对行为良好的用户进行激励，即完成目标行为次数到达一定标准则奖励一定量的"虚拟币"（通证），并将所有用户的奖惩数据同步给链中各个节点。

存在于 ESTE 中的"虚拟币"将会对应多种利润与权益（多种后援增强物），与电力市场中的期货市场一样，"虚拟币"可以与金钱进行一定汇率的等价交换，这为社会资本介入能源系统提供了可能。现阶段的新能源补贴政策为国家带来了较大的财政负担，许多地方的新能源补贴力度已经大幅下降，许多靠补贴支撑的新能源企业也纷纷倒闭，通过"虚拟币"的手段让社会资本加入到能源系统投资中，从而为新能源企业的发展提供足够的支撑。

另外，"虚拟币"还可以用来购买一些权益。根据区块链的特性，用户和发电企业发起智能合约之后，会在链中进行广播并等待执行，通常来说每个节点对链中存在的交易进行随机选取并打包，当被打包的智能合约数量达到一个设定的阈值，则形成一个区块并同步到全网所有节点，从而被打包的智能合约生效，没有被打包的智能合约只能等待下一个区块的生成。在这个过程中，区块链中存在有专门的排序节点，对智能合约的打包顺序进行排序，若是哪份智能合约提供较高的手续费（通证），它就会被排序节点放在靠前的位置。利用这一机制，利用"虚拟币"附加手续费的方法，发电企业可以获得优先交易的权益，用户也可以获得优先用电的权益。

另外，为了防止发电企业或用户大量囤积"虚拟币"造成系统体系瘫痪，设置发电企业和用户的"虚拟账户上限"，既可以加快"虚拟币"的流通，又可以减少外来资本介入扰乱系统。

## 3.4 通证经济体系应用示例——以综合能源系统为例

为了更好地描述通证经济体系在能源系统的运作原理，在综合能源系统中建立一个通证经济体系，区域综合能源系统的结构图如图 3 所示。

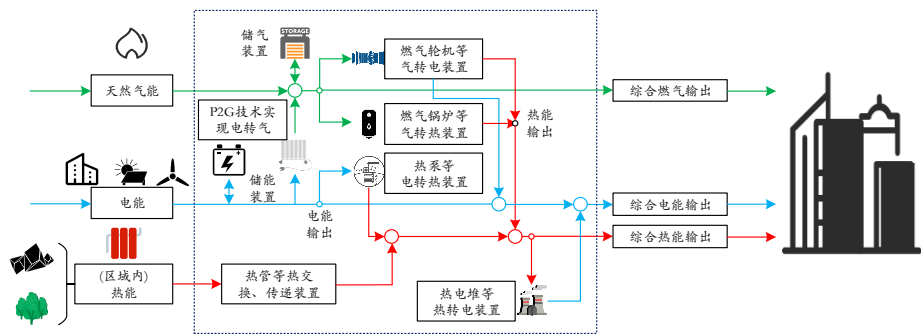

图 3 区域综合能源系统结构图

从图 3 中可以看出，综合能源系统存在着大量的电能-热能-氢能交换，最后输出的电能难以追溯，无法判断用户使用的能源类型。在区域综合能源系统中搭建区块链，为各类用户和发电企业创建账号，设置碳排放贡献因子，通过智能合约进行能源交易，一旦智能合约满足能源系统中的触发条件，区块链系统将自动执行合约，并将此消息进行全网广播，发电企业所在节点收到反馈将会对用户进行供能。值得注意的是，不满足约束条件的智能合约将会被拒绝，交易也会自动取消，并通知交易双方。其中，分析区块链会计算发电企业和用户的碳排放贡献因子，计算公式分别如式(1)和式(2)所示。

$$F_{supply} = \alpha \frac{S_{permit} - S}{S_{permit}} \quad (1)$$

$$F_{demand} = \beta \frac{P_{clean}^{electricity}}{P^{electricity}} + \gamma \frac{P_{clean}^{hot}}{P^{hot}} + \sigma f(P_{load}) \quad (2)$$

其中，$F_{supply}$ 和 $F_{demand}$ 分别表示供给侧和需求侧的碳排放贡献因子；$\alpha$、$\beta$、$\gamma$ 和 $\sigma$ 分别表示碳排放贡献因子转换系数；$S_{permit}$ 为当月发电企业允许的碳排放量标准，$S$ 为当月发电企业实际的碳排放量，其中清洁能源发电的碳排放量为 0；$P_{clean}^{electricity}$ 和 $P_{clean}^{hot}$ 分别为当月用户使用清洁能源供电和供热的总功率，$P^{electricity}$ 和 $P^{hot}$ 分别为当月用户供电和供热的总功率，$P_{load}$ 为用户提供需求侧响应的负荷功率，$f(\Box)$ 为需求侧响应功率与碳排放减少量的函数。

在支付区块链中，通证经济系统会根据各个发电企业和用户的贡献因子分别发放一定数量的通证。为了保证节能减排的效果，必须设定一个下阈值，以防止部分用户和发电企业进行投机行为，同时为了限制链中角色获得通证的数量，必须设定一个上阈值。因此，发电企业与用户获得通证的计算公式如(3)所示。

$$N_{token} = \begin{cases} \left[-\vartheta\left(e^{-F}-1\right)\right] & F < 0 \\ 0 & 0 \leq F < F_1 \\ \left[\xi\left(e^F-1\right)\right] & F_1 \leq F < F_2 \\ N_{\max} & F > F_2 \end{cases} \quad (3)$$

其中，$N_{token}$ 为可以获得的通证数量；$[\ ]$ 为取整函数；$F$、$F_1$、$F_2$ 分别为链中各主体的碳排放贡献因子以及可获得通证的碳排放贡献因子的下阈值和上阈值；$\vartheta$，$\xi$ 分别为碳排放贡献因子的惩罚系数和奖励系数；$N_{\max}$ 为链中各主体每月可以获得通证数量的上限值。值得注意的是，若主体可以获得的通证数量为负数，则需要缴纳相应数量的通证或与通证价值对应的罚款。

基于通证经济体系的区块链交易的流程图如图 4 所示。

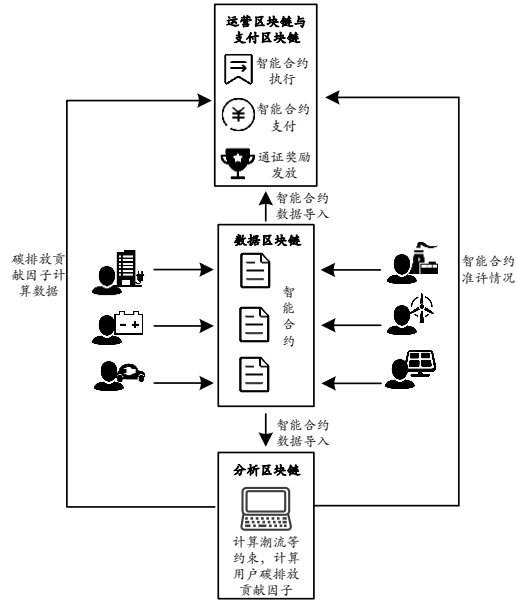

图 4 区域能源区块链交易流程

## 4 结论

未来的能源系统将不再仅限于终端能源系统，将对能源的生产、运输、分配、消费进行统一管理，具有系统庞大、管理困难、能流复杂等特点，现存的管理模式与技术将

不再适用。本文提出了在区块链上运行的通证经济学的概念、体系和技术，旨在为能源系统的管理、运营与支付等方面提供一套可行方案，旨在为解决能源系统管理困难以及弃风弃光等问题提供一条新的技术路线，从而进一步增强了能源系统的核心理念：提高能源利用率、促进清洁能源供电和鼓励用户进行需求侧响应。通证经济体系通过通证的增强作用来激励用户维持良好行为，其增强作用具有很强的适应性。另外，在通证经济体系所处的运营区块链和支付区块链上，存在与实际物理拓扑结构以及物理实体一一对应的虚拟系统结构，能在结合系统实际情况的基础上进行智能合约的交互，为通证经济体系的实际应用提供了强有力的支持。

本文所提模型从通证经济的概念入手，描述了通证经济在综合能源系统中的应用模式，未来运用人工智能技术会给出更为具体数学模型与应用流程，对涉及的许多电力系统关键技术进行详细考量、建立实际的模型与步骤，并基于实例进行验证。

## 参考文献

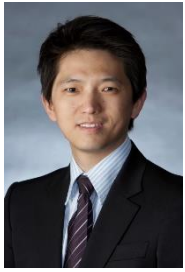

**张俊** 武汉大学电气工程学院教授. 2003 年和 2005 年分别获得华中科技大学电气工程系学士与硕士学位. 2008 年获得亚丽桑那州立大学电气工程博士学位. 主要研究方向为智能系统，人工智能，知识自动化，及其在智能电力和能源系统中的应用. 本文通信作者.

E-mail: jun.zhang@qaii.ac.cn

(ZHANG Jun Professor at School of Electrical Engineering, Wuhan University. He received his B. E. and M. E. degrees in Electrical Engineering from Huazhong University of Science and Technology, Wuhan, China, in 2003 and 2005, respectively, and his Ph. D. in Electrical Engineering from Arizona State University, USA, in 2008. His research interest covers intelligent systems, artificial intelligence, knowledge automation, and their applications in intelligent power and energy systems. Corresponding author of this paper.)

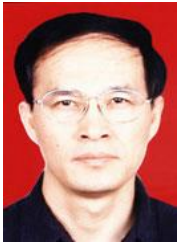

**王飞跃** 中国科学院自动化研究所复杂系统管理与控制国家重点实验室研究员. 国防科技大学军事计算实验与平行系统技术研究中心主任. 主要研究方向为智能系统和复杂系统的建模、分析与控制.

E-mail: feiyue.wang@ia.ac.cn

(WANG Fei-Yue Professor at The State Key Laboratory of Management and Control for Complex Systems, Institute of Automation, Chinese Academy of Sciences. Director of the Research Center for Computational Experiments and Parallel Systems Technology, National University of Defense Technology. His research interest covers modeling, analysis, and control of intelligent systems and complex systems.)

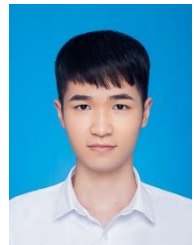

**陈思远** 武汉大学电气工程学院博士研究生. 2018 年获得武汉大学电气工程学院硕士学位。主要研究方向为智能电网，电力市场，能源系统.

E-mail: wddqcsy@whu.edu.cn

(CHEN Si-Yuan Ph.D. candidate in School of Electrical Engineering, Wuhan University. His research interest covers include smart grid, electricity market, energy system.)